# IS THE EUROPEAN MONETARY SYSTEM CONVERGING TO INTEGRATION?

Loet Leydesdorff

University of Amsterdam

Department of Science & Technology Dynamics

Nieuwe Achtergracht 166

1018 WV  Amsterdam, The Netherlands

&

Nienke A. Oomes

University of Wisconsin

Department of Economics

1180 Observatory Drive

Madison WI 53706, USA

> *So much of barbarism... still remains in the transactions*
>
> *of most civilised nations, that almost all independent countries*
>
> *choose to assert their nationality by having, to their own inconvenience*
>
> *and that of their neighbours, a peculiar currency of their own.*
>
> -- J.S. Mill (1848)

**Abstract**
The emerging system at the European level can be conceptualized as a pattern of relations among member states that tends to be reproduced despite disturbances in individual trajectories. The Markov property is used as an indicator of systemness in the distribution. The individual trajectories of nations participating in the European Monetary System is assessed using an information theoretical model that is consistent with the Markov property in the multivariate case. Economic and monetary integration are analyzed using independent data sets. Increasing integration can be retrieved in both of these dimensions, notably since the currency crises of 1992 and 1993. However, the dynamics for countries which have strongly coupled their currency to the German Mark are different from those which did not. Additionally, developments in inflation and exchange rates at the European level are assessed in relation to global developments.

1. **Introduction**

Since its ratification in December 1991, European politics has been under the spell of the Maastricht *Treaty on European Union* and, in particular, of its Protocol on the Convergence Criteria.[i] Since a sufficient number of countries were able to satisfy the criteria in 1997, the transition to the final stage of Economic and Monetary Union (EMU) in 1999 can be made: the introduction of a single European currency, the *Euro*. In addition to replacing its predecessor the Ecu, which is already in use as an accounting unit for transactions between European central banks, the Euro is meant to replace existing European currencies, by 2002 at the latest, not merely as a unit of account but as a means of payment and store of value. This replacement will necessitate a monetary policy elaborated at the level of the European Central Bank.

The introduction of a single European currency implies a transition to a system of fixed exchange rates among the participating nation states. By thus forming an economic and monetary union, European countries are expected to enjoy the benefits of lower transaction costs and greater market integration within their union.  Externally, a stronger position can be achieved with respect to non-EMU economies such as the USA.  However, the member states will lose their freedom to conduct an independent monetary policy and, in particular, they will lose the exchange rate instrument as a means to adjust for shocks and diverging developments in the real economy.

This potential cost of monetary integration has led economists as well as politicians to agree on the need for articulating some sense of prior "convergence" among candidate EMU member countries.  At the political level, the governments of the EU countries have agreed, by signing the Maastricht Treaty, to make participation in the EMU dependent on the fulfillment of so-called "convergence criteria" for prices, interest rates, fiscal deficits, and exchange rates.  Economists, however, have criticized these criteria for being either too narrow or too broad, and even for being mutually inconsistent.[ii]  As De Grauwe (1996) has noted, for instance, the criteria do not take into account the need for "real convergence," which has been considered a necessary requirement for an "optimum currency area" (Mundell, 1961; cf. Eichengreen, 1991).[iii]

How can one measure whether a system is converging to integration?  The answer to this question obviously depends on one's definition of the term "convergence."  One approach is to follow Collignon's (1994: 43) suggestion and assess the degree of European monetary integration in terms of the Maastricht convergence criteria themselves, since these are in fact the entry tickets to EMU insisted upon by governments, and are therefore important indicators for financial markets.  We shall discuss this below as *the political approach.*

Another approach, which is more in line with economic theorizing, and which we will therefore term *the economic approach*, involves defining convergence in terms of so-called "purchasing power parity" (PPP).  PPP is achieved within a system of currencies when all real exchange rates are equal to one.  Whether currencies are converging to PPP can be assessed by testing whether the real exchange rates are increasingly "covariance stationary" around one.  Empirically, this turns out not to be the case for European countries (e.g., Cheung et al., 1995; Dutt and Gosh, 1995). However, this result may be biased at least in part because of the problems involved in measuring absolute price levels and comparing them between countries.

A more popular way to test for PPP is, therefore, to analyze whether a so-called "cointegrating vector" exists, that is, to test for any linear combination of prices and nominal exchange rates which is stationary (Baillie and Selover, 1987; Banerjee et al., 1993). However, as has often been overlooked, the existence of a cointegrating vector can be interpreted as evidence of "nominal convergence" only if this vector transforms price indices into absolute price levels. Since these vectors are typically not unique, they cannot usually warrant the inference.

Given these differences in the economic and political definitions of the problem, it seems more fruitful to begin by testing independently the two questions of whether a European economic and/or monetary system is emerging. In terms of data, we will follow the "political approach" by using *nominal* exchange rates between national currencies for testing emerging systemness at the European level in the sense of the Maastricht Treaty. Real Effective Exchange Rates as constructed by the IMF will be used for testing economic integration among the nation states involved in terms of "purchasing power parity." Our results will then enable us to specify which countries are coevolving into a single union and in which respects.

Our methodology is based on Theil's (1966 and 1972) elaboration of Shannon's (1948) measure for probabilistic entropy. From this perspective, a system is considered as a pattern in the distribution that is maintained despite disturbances (Theil, 1972; Leydesdorff, 1991 and 1995). This methodology enables us to test for systemness in the data in terms of the so-called Markov property, as against non-systemness as the alternative hypothesis of trajectories at the national levels; and it also allows us to indicate sub-groups of nations that are increasingly forming a system. Before addressing the technicalities of our methods, however, let us first examine the "convergence" criteria of the Maastricht Treaty in more detail, using descriptive statistics.

2. **Convergence According to "Maastricht"**

A number of conditions for the EMU were specified in the Protocol attached to the Maastricht Treaty. As noted, these conditions are almost exclusively of a monetary kind. For example, a country will have met the "convergence criterion" for exchange rates, if it "has respected the normal fluctuation margins provided for by the ERM of the EMS without severe tensions for at least the last two years before the examination" (Protocol on the Convergence Criteria, at p. 185).

The "ERM of the EMS" is the Exchange Rate Mechanism (ERM) of the European Monetary System (EMS). It was established in 1979 with the purpose of stabilizing exchange rates between the currencies of member states of the European Union (EU).[iv] This was done by keeping the nominal market exchange rate of each ERM currency vis-à-vis any other ERM currency within preassigned margins around bilateral central parities. These bilateral fluctuation margins, which served as mandatory intervention limits, were set, for most countries, at 2.25 percent on each side of the central parity, so that the width of their fluctuation bands was a total of 4.5 percent.

Since the currency crisis of August 1993, however, the margins for all ERM currencies, except for the German Mark and the Dutch Guilder, have been widened to 15 percent of the central rate, so that the fluctuation band width is now 30 percent. This increase has obviously made it easier for countries to respect the bounds, as well as to satisfy the Treaty's additional requirement that "the member state shall not have devalued its currency's bilateral central rate against any other member state's currency on its own initiative for the same period."

Although realignments of central rates did occasionally happen in the past, no changes in parity grids have taken place since March 1995, when the Spanish Peseta was devalued by 7 percent and the Portuguese Escudo by 3.5 percent. Thus, according to the Treaty's convergence criterion, nominal exchange rates have sufficiently "converged" by now in order to allow for their definite fixation by introducing the Euro. As noted, however, this is mainly a political convergence criterion, which is not particularly supported by economic theories of exchange rate determination. According to the latter, movements in nominal exchange rates should reflect those in relative price levels.

The Maastricht Treaty indeed specifies as an additional convergence criterion the requirement that EMU countries have "[a] price performance that is sustainable and an average rate of inflation, observed over a period of one year before the examination, that does not exceed by more than 1.5 percentage points that of, at most, the three best performing member states in terms of price stability. Inflation shall be measured by means of the consumer price index on a comparable basis, taking into account differences in national definitions." The Treaty does not specify how exactly differences in national definitions should be taken into account. Yet the "raw" annual data on consumer price indices (CPIs) provided by the IMF do suggest "convergence" in the inflation rates of the twelve EU countries that signed the Maastricht Treaty (see Figure 1). Based on these data, the convergence criterion for price stability was met by ten countries in 1996.

**Figure 1. Inflation Rates 1980-1996**

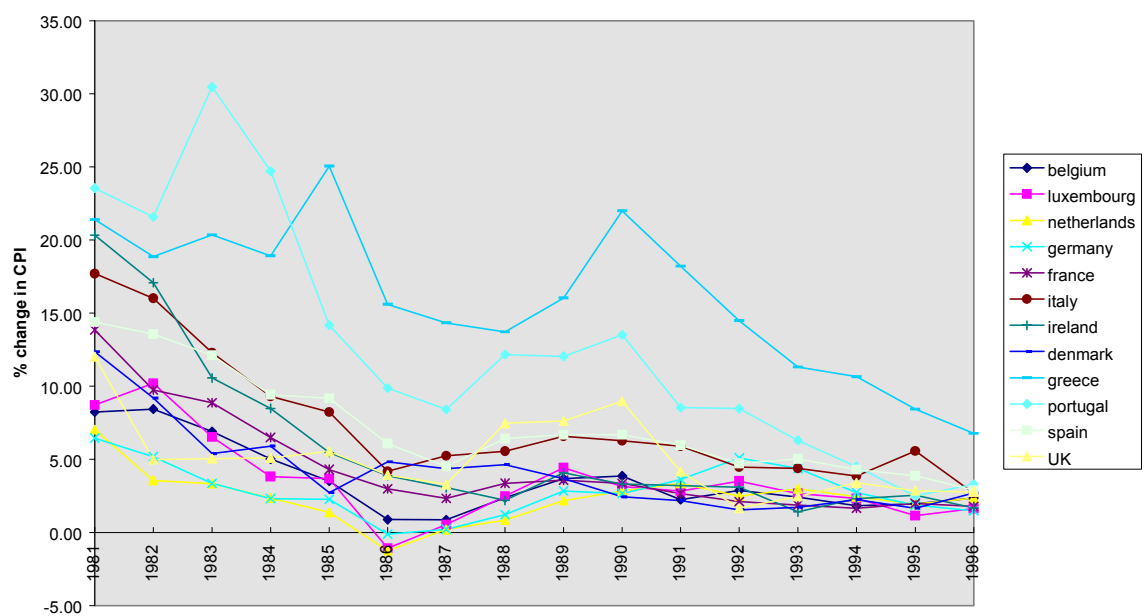

This measurement of "convergence" in terms of the growing number of countries whose inflation rates exceed the mean of the three lowest inflation rates in any period by not more than 1.5 percent, is completely *ad hoc*. One does not need to run a statistical test in order to see, simply by looking at Figure 1, that the relative distance between inflation rates has decreased over time. Apart from Greece, all these EU countries now have inflation levels below 5 percent. However, this low level of inflation need not have been anticipated by the Maastricht Treaty. According to its political definition, "convergence" could have occurred just as well at much higher levels of inflation. Furthermore, convergence in the rate of change of prices does not imply convergence in price levels. In fact, if price levels were different to begin with, convergence in inflation rates would have reinforced these differences.

We propose to move to alternative measures of convergence, which are informed by systems theory, and which enable us to estimate convergence in price levels and nominal exchange rates independently. The implications for monetary and economic integration will be discussed in the concluding section. We will be able to show that the two types of integration are both taking place, although they have hitherto not been coupled.

3. **Methods**

Is a European system emerging on top of a set of firmly integrated national systems? In order to answer this question, let us define a second-order system as a specific distribution or "network" of exchange relations which tends to be reproduced over time. The national systems can then be considered as the "nodes" of the European network, and the question to be answered is whether structure is maintained at the network level inspite of disturbances at the nodes. The emerging network is a system insofar as it is reproduced as a pattern that is relatively independent of the historical trajectories of the composing (sub-)systems. Note that the emerging system is thus not considered an aggregate, but as an interactive effect of the relations among national systems.

In other words, the nation states follow their own historical trajectories as economic and monetary systems, while the European system unfolds on top of them as an emerging "exchange rate regime." Since a system is expected to stabilize the variation at its nodes, the current state of a system then provides the best expectation of its next state. When the current state is, in fact, the sole determinant of the next state, the system can be said to satisfy the Markov property.[v]

The "states" of a system need not be one-dimensional. The system as a whole is, typically, a multivariate arrangement that is reproduced at each moment in time, while the underlying nodes develop over time as univariate trend lines. Information theory allows us to separate these "micro" and "macro" developments analytically, and then to test for systemness at the level of the distribution. This is done by comparing the predictions based on the assumption of systemness with those obtained on the basis of the individual time series autoregression. The difference between these two expectations can be quantified in terms of Shannon's (1948) information measure.

3.1     *Information Theory*

The information measure developed by Shannon (1948) is based on the assumption that the amount of information conveyed by the occurrence of an event *i*, or by the message stating that this event has occurred, is inversely proportionate to the probability $p_i$ of that event occurring. Intuitively, when *i* is expected to take place with high probability, the fact that it did indeed take place contains very little information. Conversely, when the occurence of *i* is improbable, the message stating this occurrence must be considered highly informative.

Using the negative of the logarithm as a particularly convenient function which is continuous, monotonically decreasing in $p_i$ and non-negative on the domain [0,1], Shannon (1948) defined the amount of information as:

$$h(p_i) = -\,^2\!\log(p_i)$$

where 2 was chosen as the base of the logarithm so as to let information be measured in binary digits (bits).[vi] Using this definition, the *expected* amount of information over the entire distribution of events turned out to be equivalent to what is also called "probabilistic entropy" because of the formal similarity with the entropy function in thermodynamics:

$$H_i = - \Sigma_i \, p_i \, {}^2\log p_i$$

The dynamic equivalent of this measure was provided by Theil (1972), as the expected information *I* of the message that the *prior* distribution ($\Sigma \, p_i$) has been transformed into the *posterior* distribution ($\Sigma \, q_i$):[vii]

$$I_i = \Sigma_i \, q_i \, {}^2\log q_i/p_i$$

The extension of both formulas to the multivariate case is straightforward:

$$H_{ijk\ldots} = - \Sigma_{ijk\ldots} \, p_{ijk\ldots} \, {}^2\log p_{ijk\ldots}$$
$$I_{ijk\ldots} = \Sigma_{ijk\ldots} \, q_{ijk\ldots} \, {}^2\log q_{ijk\ldots}/p_{ijk\ldots}$$

3.2     *Application of Shannon's measure to applied forecasting*

The Shannon/Theil methodology enables us to model *change* in univariate and multivariate distributions within a single and integrated framework. Although Theil (1966, Chapter 11) suggested an extension of the dynamic algorithm to the case of time series analysis, he himself never related this to the Markov chain perspective for systems (cf. Theil, 1972, Chapter 5). In *Appendix 1*, we therefore derive an algorithm that is consistent with the Markov property for the multivariate case, but which allows us, in addition, to make best predictions for the next year based on autoregression in the univariate trend lines of sub-systems. Thus, we will be able to compare the assumption of separate trajectories for each country with the assumption of systemness in the exchange relations in terms of the quality of the prediction.

The basic idea is that one is able to compare the distribution of a given time-series (1, 2, ...., n) with a hypothesized distribution, by comparing the prediction for (n + 1) with its actual value. Given that *no other* information is available, the expectation is that $I = 0$. It can be shown (see the Appendix) that the expected value $F_{n+1}$ can then be written as:

$$F_{n+1} = F_n \{(\Sigma_1^n F_i - F_1)\} / \{(\Sigma_1^n F_i - F_n)\}$$

Note that this formula can be applied to any series of discrete values which represents an individual history. It can additionally be shown that in the case of a multivariate set, this formula is consistent with the noted Markov property.

For example, on the basis of Table 1 (which constitutes one of the datasets to be explained in more detail below) one is able to make a prediction of the (n+1)th observation, based either on the rows representing the histories of each national currency or the columns representing the distribution of the system of currencies in each year. Since the rows and columns are orthogonal, these two predictions are completely independent, yet they can both be expressed in bits of information and can therefore be compared (after proper normalization).

Given the two prior predictions ($\Sigma\ p_i$ and $\Sigma\ p_i'$) on the one hand, and the actually occurring posterior distribution $\Sigma\ q_i$ on the other, the improvement of the revised prediction $\Sigma\ p_i'$ over the original prediction based on $\Sigma\ p_i$ is defined as follows:

$$I_i = \Sigma_i\ q_i \log (q_i/p_i) - \Sigma_i\ q_i \log (q_i/p_i')$$

(= original prediction - revised prediction)

$$= \Sigma_i\ q_i \log (p_i'/p_i)$$

In other words: whether the revision of the prediction implies an improvement of the prediction depends on the relative uncertainties associated with the two predictions. If the revised prediction is associated with a smaller uncertainty than the original one, $I$ is positive and the prediction is improved; negative values of $I$ indicate a worsening of the prediction.

Since we are not testing a single hypothesis against the data, but two hypotheses against each other using limited data, "significance" does not have its common statistical meaning (cf. McCloskey, 1985). For the reader's information, we will place results between brackets when the differences between the two predictions are less than an order of magnitude, that is, if $I$ on the basis of the multivariate predictions is less than or larger than ten times the summation of univariate predictions (after proper normalization). Since the expected information of the improvement is additive, hence decomposable, this methodology enables us to evaluate for which (group of) countries the assumption of systemness provides a better prediction than than the alternative assumption of independent development (cf. Theil, 1972).

The qualitiy of the univariate predictions is, of course, sensitive to the length of the time series on which they are based. We exhaust our data by comparing all possible, yet consecutive time series with a minimum of three years including the year $m$ under study (that is $m$, $m$-$1$, and $m$-$2$). The comparison with the multivariate prediction is based on the lowest $I$ for each individual country. As the time series are longer for more recent years, the univariate predictions are increasingly precise. Consequently, the test for systemness at the network level in recent years is more powerful. This is relevant with respect to our research question of whether a European system has recently emerged.

4. **Data**

While nominal exchange rates denote how many units of a foreign *currency* can be obtained in exchange for one unit of a domestic *currency*, real exchange rates can be considered as a measure of how many foreign *goods* can be obtained in exchange for one unit a domestic *goods*.

4.1    *Nominal Exchange Rates*

We used the nominal exchange rates data provided by the *OECD Economic Outlook* of December 1996 (Annex Table 37). This "OECD data set" is exhibited in Table 1. The table contains data for 31 national currencies vis-à-vis the US dollar, based on averages of daily rates for the period 1984 to 1995.[viii] Additionally, estimates and projections are given for 1996, 1997, and 1998. Since the data were estimated on 4 November 1996, we only used the values for 1996.[ix]

For the years 1984-1991, data are missing for the Czech Republic, Hungary, and Poland. Given that our research question did not involve these three countries in particular, we decided to leave them out of consideration, hence reducing the number of currencies to 28. In order to make the analysis based on nominal exchange rates comparable with that based on real effective exchange rates (REER), we had to change the unit of account from the US Dollar to the German Mark and the base year to 1990. These normalizations are strictly technical.

4.2    *Real Effective Exchange Rates*

Given that no data on absolute price levels are available, the data which best approximate real exchange rates, in our view, are the Real Effective Exchange Rate (REER) Indices provided by the IMF. The REER indices are constructed by using trade weights, which take into account the relative importance of a country's trading partners in its direct bilateral relations with them in both home and foreign markets, the competitive relations with third countries in particular markets, and the differences among countries in the importance of foreign trade to the manufacturing sector. In other words, the indices adjust for movements in costs and prices affecting exports and imports of manufactured goods.[x]

In order to make our REER based analysis comparable to the analysis based on nominal exchange rates, we constructed annual REER indices by averaging over each year's monthly values (see Table 2). We limited the comparison to the twelve EC countries that signed the Maastricht Treaty only, for reasons which will become clear below. Since the IMF (1997) does not provide REER indices for Greece, this country had to be left out of consideration. (See, however, the remarks in notes 4 and 9 about the special status of Greece given this country's monetary policies.)

5    **The global system of currencies**

5.1   *Descriptive statistics using exploratory factor analysis*

Using the annual nominal exchange rates of Table 1, factor analysis teaches us that during this period two monetary systems can be distinguished. A two-factor solution (accounting for 87.5% of the variance, using Varimax) is revealed in Table 3. On the one hand, one observes a group of currencies (Factor Two) that correlates with the US Dollar, and on the other, a group that correlates either positively or negatively with the German Mark and the Japanese Yen (Factor One). By sorting factor loadings within Factor One, the prevailing structure of the European bloc during this period is clearly made visible. Relatively floating currencies are distinguished from the ones which have deliberately been anchored to the German Mark. In this representation both the French Franc and the Japanese Yen are represented within this latter group. The UK Sterling, on the other hand, exhibits a pattern similar to the Italian Lire, and, for example, the Mexican Peso. These currencies exhibit a negative correlation to the DM bloc, but only some of them join the UK Sterling in also correlating with Factor Two representing the US Dollar bloc.

The best representative of the US Dollar bloc during this period (1984-1996) has been the Hong Kong Dollar.[xi] Note Singapore's secondary orientation towards the Japanese Yen. Furthermore, the French Franc is the only member of the group correlated with the German Mark and with nevertheless a positive loading on the factor representing the US Dollar bloc. In summary, the dynamics *within* the European system seem to prevail as a pattern at the level of the global system of currencies. One legitimate question, then, is under what conditions would a unified Europe be able to profit from stabilizing its exchange rate regime?

5.2    *The Markov test in the case of the global system*

Let us first consider the null hypothesis that the 28 currencies of the OECD data set constitute a single system, in the sense of exhibiting the Markov property. Using the methods described in section three, we will test this hypothesis by comparing, for each year, the univariate forecasts with the multivariate forecast.

Table 4 gives an overview of the *improvement* of the predictions based on the assumption of systemness (i.e., multivariate forecasts) over the predictions based on the alternative hypothesis (i.e., the normalized sum of univariate forecasts) for the various years. If the improvement is positive, the hypothesis of systemness is corroborated, while a negative value can be considered as a rejection of this hypothesis. The first column displays the result on the basis of using the raw data. As it turns out, the systemness assumption is *always rejected*. However, this result is potentially flawed, since the distribution of exchange rates is, in general, not invariant with respect to changes over time in the units in which currencies are denominated.[xii]

The second column of Table 4 is based on normalization by setting 1990 = 100 for each country and using the German Mark. The results given in this column imply that the assumption of a "global system" can then be rejected in almost all years.[xiii]

6. **Is a European System Emerging?**

6.1    *European Union (EU) versus European Community (EC)*

Let us now limit our analysis to the 15 member states of the current European Union (EU). Since Belgium and Luxembourg use a single currency, this implies a set of 14 currencies. As a second frame of reference, we will use the 12 member countries of the former European Community (EC) which originally signed the Maastricht Treaty. As explained above, Greece is removed from this comparison for various reasons. Hence, our second frame of reference will be a set of 11 EC countries.

Table 5 provides the results for these two European systems of reference, EU and EC, in a format similar to Table 4. On the basis of nominal exchange rates, the group of 11 EC members appears to have functioned as a single system, with the exception of some disruptions in the early 1990s. As noted before, this is likely to indicate the currency crises of 1992 and 1993. For the EU group, however, the assumption of systemness almost never leads to an improvement of the prediction, although it seems to do slightly better after 1993.

The third column of Table 5 provides a similar test for the real exchange rates for the 11 EC member states. (Belgium and Luxembourg can now be taken separately, since they differ in terms of prices levels and trade patterns.) The hypothesis of systemness is confirmed, in the sense that the Markov assumption cannot be rejected in most of the years, especially in the most recent years.

### 6.2   *Decomposition of the European "System"*

Since the improvement of the prediction is based on a summation ($I = \Sigma_i \, q_i^{\,2} \log (p'_i / p_i)$), one is able to raise the question of which countries are contributing most to the stability of the European system in which year. Table 6 gives the decomposition for the eleven countries involved in the Maastricht Treaty. Note that, for reasons of normalization, one can only compare the decomposition over the years in terms relative to the (sum-)value of the set (that is, the "EC"). Since sumvalues are positive for most of the years (see also Table 5) a negative sign can be used as a first order indicator of a country's failure to integrate.

On the basis of Table 6, it is clear that monetary integration does not imply economic integration, and *vice versa*. While the ERM crises of the early 1990s are clearly visible in the case of currencies like the UK Sterling, the Italian Lire, and the Spanish Peseta, their underlying economies seem to have become more, rather than less, integrated with the rest of the European economic system. Actually, for Spain and the UK the assumption of the Markov property holds since 1992 in terms of real exchange rates.

On the other hand, a number of countries which were coupled to Germany in terms of montary policies, appear to have been less successful in keeping step with the German economy. When real effective economic exchange rates are used instead of nominal exchange rates, predictions based on the Markov property fail the test since 1993 in the case of Belgium, Denmark, France, and The Netherlands. Note that this does not imply, for instance, that Spain is doing better than France in terms of economic performance. What it means is that the economies of these countries are differently coupled in terms of their co-evolution with an emerging European system.

# 7	Summary and Conclusions

The main research question of this study has been whether a European system is emerging in the monetary and the economic dimension. A first conclusion is that, according to the political approach most clearly represented by the Maastricht Treaty, the European *Monetary* System certainly is "converging" to integration. Measured by the standards of the Protocol on the Convergence Criteria, the European system has indeed "converged" with respect to inflation rates, interest rates, fiscal deficits, as well as in terms of nominal exchange rates. The introduction of a single currency, which effectively implies a system of fixed nominal exchange rates, can be considered as the "end" of this monetary integration process.

The European Monetary System was established in 1979 with the explicit purpose of attaining this end, by narrowing the "band widths" within which the exchange rates of its members would be allowed to fluctuate. In 1992-1993, this so-called Exchange Rate Mechanism (ERM) proved too rigid, when serious currency crises forced the UK and Italy to withdraw from the system. Eventually, the crises led to a substantial widening of the band widths in the fall of 1993. Although this new arrangement eroded the ERM in practical terms, it became considerably easier for countries to meet the "convergence criteria" for exchange rates laid down in the Maastricht Treaty, according to which eligible EMU members must have respected the normal fluctuation margins of the ERM "without severe tensions for at least the last two years before the examination."

On the basis of information theory, we have shown that, if "convergence" is defined in terms of contributions to system formation, then those countries which adopted a floating or strongly fluctuating nominal exchange rate regime after the currency crisis, have failed to converge in nominal terms. In terms of "real convergence," however, these countries turn out to have performed "better," in the sense that they have contributed more to the formation of a system at the economic level than countries whose monetary regimes have been coupled strongly with that of Germany. In this case, monetary integration has obviously not implied economic integration, or *vice versa*. Our results suggest that strong monetary coupling has provided a negative feedback to economic integration in more recent years.

In our opinion, more attention should have been paid to the theory of "optimum currency areas" (Mundell, 1961). From this perspective, a successful monetary union requires a considerable amount of real wage flexibility and labor mobility in order to compensate for "real" divergence in terms of output and employment rates. Heylen et al. (1995) have shown that current rigidities in European labour markets make it unlikely that these conditions can be met. The authors of the Maastricht Treaty, on the other hand, seem to have presumed that the European Union constitutes a so-called "efficient market" in which prices reflect all relevant information, so that monetary integration will imply economic integration. On the basis of our results, we conclude that this assumption is mistaken.

* The authors acknowledge partial funding by the European Commission, TSER project PL97-1296, entitled "The Self-Organization of the European Information Society."

**Notes**

i. Council of the European Communities, and Commission of the European Communities (1992), *Treaty on European Union,* Luxembourg: Office for Offical Publications of the European Communities

ii. For the purpose of reaching nominal convergence alone the criteria seem partially redundant. Long-term interest rates, for instance, are generally considered to reflect (expected) inflation differentials, as well as changes in nominal exchange rates and fiscal deficits. Since all these variables depend, in turn, on monetary policy, it has been argued that "monetary convergence is not really a pre-condition for monetary union, but is just an obvious effect of it." (Giovannini 1994: 189). Bofinger (1994) has argued that the very expectation of future exchange rate fixing causes nominal variables to converge well before the actual monetary union is established.

iii. On the relation between nominal and real convergence, see, e.g., Heylen *et al*. (1995).

iv. Starting in 1979 with the Benelux, Denmark, France, Germany, and Italy, the EMS was later joined by the UK, Ireland, Spain, Portugal, Finland, Sweden, and Austria, and includes now all fifteen EU member states, most of which have also participated in the ERM. During the EMS crisis in the fall of 1992, the UK, which had joined in 1990, was forced to withdraw from the ERM, along with Italy. Since 1992, Finland, Italy, Portugal, Sweden, and the UK have maintained floating exchange relations. The Greek Drachme, however, does not participate in the ERM mechanism.

v. This definition can be extended to a so-called second- or higher-order Markov property in which more previous states are taken into account. In general, however, a system's development is not dependent on the historical trajectories of its elements.

vi. In terms of bits, the occurrence of a 50-50 event produces exactly one unit of information: $h(0.5) = {}^2\log(2)$.

vii. It can be shown (Theil 1972: 59f.) that $I >= 0$, that is, a change in the distribution is always associated with positive information being generated unless the distribution remains the same.  In the latter case, $I$ is equal to zero because no (probabilistic) entropy is generated when there is no change in a system.

viii. The OECD additionally provides rates for the European Currency Unit ECU.  However, we have left these normalized values out of consideration in this study since it would beg the research question.  (SDR at the bottom line of this table stands for "Special Drawing Rights" of the IMF.  This information, however, is not relevant in the context of this study.)

ix. The OECD notes that the projections are based on the technical assumption that exchange rates remain at their levels of 4 November 1996, except for Greece, Hungary, Poland, and Turkey, where exchange rates vary according to official exchange rate policy.

x. The REER indices are provided by the IMF in terms of both relative consumer price indices (CPI) and unit labor costs (ULC).  These are two marginally different measures of domestic prices and costs (IMF 1981).  In this study, we shall use CPI-based REER indices since our interest is primarily in terms of purchasing power parity.

xi. As perhaps expected, a one-factor solution (explaining 73.8% of the variance) exhibits the DM-block consisting of the Netherlands (-0.99555), Austria (-0.99541), and Germany (-0.99514) as the principal component, while the UK (0.98867) and Greece (0.96876) are the most pronounced representatives of the alternative.  The US and Hong Kong Dollars follow with factor loading of 0.84205 and 0.83210, respectively.  The Yen is again linked with the DM, as is the French Franc, yet at lower levels (-0.83619 and -0.59192, respectively).

xii. The factor analysis (used above) is normalized for size effects because of the Pearson correlations underlying it as a first step.

xiii. In some years, however, the assumption of a global monetary system provided the better prediction using this test.  This is in line with our expectation of the possibility of weak systemness at the global level because of the ongoing operation of markets.

**Table 1:** Nominal exchange rates vis-à-vis the US dollar (the OECD data set).
Source: OECD, 1996.

| | 1984 | 1985 | 1986 | 1987 | 1988 | 1989 | 199 | 1991 | 1992 | 1993 | 1994 | 1995 | 1996* | 1997* | 1998* |
|---|---|---|---|---|---|---|---|---|---|---|---|---|---|---|---|
| US | 1.0 | 1.0 | 1.0 | 1.0 | 1.0 | 1.0 | 1.0 | 1.0 | 1.0 | 1.0 | 1.0 | 1.0 | 1.0 | 1.0 | 1.00 |
| JAPAN | 237.6 | 238.6 | 168.5 | 144.6 | 128.1 | 138.0 | 144.8 | 134.5 | 126.7 | 111.2 | 102.2 | 94.1 | 109.0 | 114.1 | 114.1 |
| GERMANY | 2.846 | 2.944 | 2.173 | 1.797 | 1.756 | 1.880 | 1.616 | 1.659 | 1.562 | 1.653 | 1.623 | 1.433 | 1.502 | 1.513 | 1.513 |
| FRANCE | 8.739 | 8.984 | 6.927 | 6.009 | 5.957 | 6.380 | 5.446 | 5.641 | 5.294 | 5.662 | 5.552 | 4.991 | 5.105 | 5.118 | 5.118 |
| ITALY | 1757 | 1909 | 1491 | 1297 | 1302 | 1372 | 1198 | 1241 | 1232 | 1572 | 1613 | 1629 | 1543 | 1520 | 1520 |
| UK | 0.752 | 0.779 | 0.682 | 0.612 | 0.562 | 0.611 | 0.563 | 0.567 | 0.570 | 0.666 | 0.653 | 0.634 | 0.642 | 0.609 | 0.609 |
| CANADA | 1.295 | 1.366 | 1.389 | 1.326 | 1.231 | 1.184 | 1.167 | 1.146 | 1.209 | 1.290 | 1.366 | 1.372 | 1.361 | 1.337 | 1.337 |
| AUSTRALIA | 1.141 | 1.432 | 1.496 | 1.429 | 1.281 | 1.265 | 1.282 | 1.284 | 1.362 | 1.473 | 1.369 | 1.350 | 1.279 | 1.269 | 1.269 |
| AUSTRIA | 20.01 | 20.69 | 15.27 | 12.64 | 12.34 | 13.23 | 11.37 | 11.67 | 10.99 | 11.63 | 11.42 | 10.08 | 10.56 | 10.64 | 10.64 |
| BELGIUM/LUX | 57.76 | 59.43 | 44.69 | 37.34 | 36.77 | 39.4 | 33.42 | 34.16 | 32.15 | 34.55 | 33.46 | 29.5 | 30.91 | 31.21 | 31.21 |
| CZECH/SLO | *** | *** | *** | *** | *** | *** | *** | 29.47 | 28.26 | 29.15 | 28.79 | 26.54 | 27.13 | 26.96 | 26.96 |
| DENMARK | 10.36 | 10.59 | 8.09 | 6.84 | 6.73 | 7.31 | 6.19 | 6.39 | 6.04 | 6.48 | 6.36 | 5.6 | 5.79 | 5.81 | 5.81 |
| FINLAND | 6.003 | 6.196 | 5.070 | 4.396 | 4.186 | 4.288 | 3.823 | 4.043 | 4.486 | 5.721 | 5.223 | 4.367 | 4.583 | 4.536 | 4.536 |
| GREECE | 112.7 | 138.1 | 139.5 | 135.2 | 141.7 | 162.1 | 158.2 | 182.1 | 190.5 | 229.1 | 242.2 | 231.6 | 241.4 | 250.1 | 252.60 |
| HUNGARY | *** | *** | *** | *** | *** | *** | *** | 74.8 | 79.0 | 91.9 | 105.1 | 125.7 | 152.7 | 176.2 | 203.40 |
| ICELAND | 31.73 | 41.54 | 41.1 | 38.68 | 43.05 | 57.11 | 58.38 | 59.1 | 57.62 | 67.64 | 69.99 | 64.77 | 66.65 | 66.38 | 66.38 |
| IRELAND | 0.923 | 0.946 | 0.747 | 0.672 | 0.657 | 0.706 | 0.605 | 0.622 | 0.588 | 0.683 | 0.670 | 0.624 | 0.627 | 0.611 | 0.611 |
| MEXICO | 0.192 | 0.327 | 0.639 | 1.418 | 2.281 | 2.495 | 2.841 | 3.022 | 3.095 | 3.115 | 3.389 | 6.421 | 7.604 | 7.905 | 7.905 |
| NETHERLANDS | 3.209 | 3.322 | 2.450 | 2.026 | 1.977 | 2.121 | 1.821 | 1.870 | 1.759 | 1.857 | 1.820 | 1.605 | 1.682 | 1.697 | 1.697 |
| NEWZEALAND | 1.767 | 2.026 | 1.917 | 1.695 | 1.529 | 1.674 | 1.678 | 1.729 | 1.860 | 1.851 | 1.687 | 1.524 | 1.454 | 1.411 | 1.411 |
| NORWAY | 8.169 | 8.594 | 7.392 | 6.737 | 6.517 | 6.903 | 6.258 | 6.484 | 6.214 | 7.094 | 7.057 | 6.337 | 6.450 | 6.370 | 6.370 |
| POLAND | *** | *** | *** | *** | *** | *** | *** | 1.058 | 1.363 | 1.814 | 2.273 | 2.425 | 2.699 | 3.047 | 3.371 |
| PORTUGAL | 146.4 | 169.9 | 148.2 | 140.8 | 143.9 | 157.1 | 142.3 | 144.4 | 134.8 | 160.7 | 166.0 | 149.9 | 153.9 | 152.9 | 152.90 |
| SPAIN | 160.8 | 170.1 | 140.0 | 123.5 | 116.5 | 118.4 | 101.9 | 103.9 | 102.4 | 127.2 | 134.0 | 124.7 | 126.4 | 127.5 | 127.50 |
| SWEDEN | 8.273 | 8.602 | 7.124 | 6.337 | 6.129 | 6.446 | 5.918 | 6.045 | 5.823 | 7.785 | 7.716 | 7.134 | 6.685 | 6.589 | 6.589 |
| SWITZERLAND | 2.350 | 2.357 | 1.798 | 1.491 | 1.463 | 1.635 | 1.389 | 1.434 | 1.406 | 1.477 | 1.367 | 1.182 | 1.231 | 1.271 | 1.271 |
| TURKEY | 363 | 520 | 669 | 855 | 1421 | 2120 | 2606 | 4169 | 6861 | 10964 | 29778 | 45738 | 81199 | 139519 | 231335 |
| TAIPEI | 39.66 | 39.85 | 37.84 | 31.92 | 28.57 | 26.28 | 26.63 | 26.55 | 25.02 | 26.35 | 26.45 | 26.49 | 27.45 | 27.5 | 27.50 |
| HONGKONG | 7.819 | 7.791 | 7.804 | 7.795 | 7.806 | 7.799 | 7.789 | 7.770 | 7.739 | 7.735 | 7.728 | 7.734 | 7.734 | 7.732 | 7.732 |
| KOREA | 806.6 | 870.9 | 881.0 | 825.0 | 730.0 | 669.2 | 708.0 | 733.2 | 780.0 | 802.4 | 804.3 | 771.4 | 803.0 | 825.3 | 825.30 |
| SINGAPORE | 2.133 | 2.200 | 2.177 | 2.106 | 2.013 | 1.950 | 1.812 | 1.727 | 1.628 | 1.615 | 1.527 | 1.417 | 1.411 | 1.408 | 1.408 |
| | | | | | | | | | | | | | | | |
| ECU | 1.272 | 1.322 | 1.019 | 0.868 | 0.846 | 0.908 | 0.788 | 0.809 | 0.773 | 0.854 | 0.843 | 0.765 | 0.787 | 0.785 | 0.785 |
| SDR | 0.972 | 0.966 | 0.853 | 0.774 | 0.742 | 0.780 | 0.738 | 0.731 | 0.710 | 0.716 | 0.699 | 0.659 | 0.689 | 0.691 | 0.691 |

* On the technical assumption that exchange rates remain at their levels of 4 November 1996, except for Greece, Hungary, Poland and Turkey where exchange rates vary according to the official exchange rate policies

**Table 2**

Real Economic Exchange Rates based on Consumer Price Indices for 11 out of 12 countries having signed the Maastricht Treaty.  (Source: IMF 1997.)

|          | 1980   | 1981   | 1982   | 1983   | 1984   | 1985   | 1986   | 1987   | 1988   | 1989   | 1990 | 1991   | 1992   | 1993   | 1994   | 1995   | 1996   |
|----------|--------|--------|--------|--------|--------|--------|--------|--------|--------|--------|------|--------|--------|--------|--------|--------|--------|
| BELGIUM  | 102.92 | 104.65 | 93.80  | 92.14  | 95.79  | 99.80  | 98.25  | 98.05  | 97.85  | 98.86  | 100  | 100.87 | 98.35  | 95.34  | 96.70  | 95.23  | 94.81  |
| DENMARK  | 87.66  | 91.39  | 88.06  | 88.25  | 90.69  | 94.52  | 94.97  | 96.11  | 98.08  | 98.58  | 100  | 98.35  | 96.02  | 93.39  | 92.78  | 91.95  | 92.65  |
| FRANCE   | 101.58 | 106.11 | 99.56  | 95.20  | 97.83  | 103.21 | 100.99 | 98.97  | 99.16  | 99.49  | 100  | 98.54  | 96.57  | 94.62  | 94.23  | 91.96  | 93.18  |
| GERMANY  | 100    | 100    | 100    | 100    | 100    | 100    | 100    | 100    | 100    | 100    | 100  | 100    | 100    | 100    | 100    | 100    | 100    |
| IRELAND  | 82.03  | 88.99  | 93.83  | 94.39  | 98.51  | 102.67 | 103.96 | 99.29  | 98.30  | 98.53  | 100  | 99.03  | 97.99  | 88     | 88.10  | 84.30  | 86.81  |
| ITALY    | 76.24  | 81.46  | 81.02  | 84.84  | 89.38  | 91.63  | 92.69  | 92.40  | 93.80  | 98.38  | 100  | 102.36 | 97.35  | 79.63  | 77.12  | 68.32  | 76.93  |
| LUXEMB.  | 99.99  | 106.57 | 98.52  | 98     | 102.27 | 105.76 | 100.68 | 97.18  | 98.96  | 101.11 | 100  | 101.92 | 98.83  | 96.06  | 96.76  | 93.68  | 93.23  |
| NETHERL. | 101.64 | 103.23 | 104.48 | 102.38 | 103.50 | 104.07 | 103.94 | 102.44 | 102.58 | 100.95 | 100  | 99.88  | 98.16  | 96.73  | 97.22  | 96.30  | 96     |
| PORTUGAL | 85.29  | 98.68  | 96.39  | 88.93  | 94.51  | 98.40  | 92.08  | 87.92  | 90.71  | 97.11  | 100  | 109.11 | 114.11 | 107.31 | 105.80 | 103.56 | 106.04 |
| SPAIN    | 82.87  | 85.46  | 83.82  | 73     | 78.53  | 82.18  | 82.68  | 83.29  | 89.28  | 97.03  | 100  | 102.96 | 98.95  | 85.43  | 81.33  | 78.67  | 81.65  |
| UK       | 100.63 | 113.45 | 107.49 | 98.09  | 98.14  | 102.93 | 90.50  | 87.50  | 96.87  | 99.65  | 100  | 104.12 | 96.90  | 84.06  | 84.08  | 77.06  | 79.31  |

**Table 3**
Factor Analysis of Nominal Economic Exchange Rates 1984-1996 as provided in Table 1 above (the OECD data set).

Varimax   Rotation  1,  Extraction  1,  Analysis  1 - Kaiser Normalization.

  Varimax converged in    3 iterations.

Rotated Factor Matrix:

|  | FACTOR 1 | FACTOR 2 |
|---|---|---|
| ITALY | .97271 | .19603 |
| SWEDEN | .92794 | .32636 |
| TURKEY | .92131 | .07906 |
| SPAIN | .90493 | .14521 |
| GREECE | .80595 | .53880 |
| UK | .79864 | .58279 |
| FINLAND | .79276 | .22311 |
| MEXICO | .73907 | .50988 |
| IRELAND | .71803 | .57752 |
| ICELAND | .66754 | .63320 |
| DENMARK | -.92685 | -.30566 |
| FRANCE | -.89369 | .22445 |
| BELGIUM/LUX | -.87618 | -.47089 |
| JAPAN | -.87368 | -.21942 |
| NETHERLANDS | -.81883 | -.56667 |
| AUSTRIA | -.80986 | -.57881 |
| GERMANY | -.80863 | -.58004 |
| SWITZERLAND | -.80463 | -.33249 |
| HONGKONG | .36416 | .91550 |
| US | .38144 | .90861 |
| AUSTRALI | .46248 | .85819 |
| NEWZEALA | .27835 | .84627 |
| SINGAPORE | -.46517 | .80755 |
| KOREA | .50386 | .79325 |
| NORWAY | .59151 | .78403 |
| CANADA | .59346 | .74605 |

```
PORTUGAL         .63454                      .71919
  TAIPEI        -.11229                      .42892
```

**Table 4**
Improvement (or worsening) of the prediction on the basis of the assumption of systemness in the data in comparison to univariate timelines, for the case of 28 worldwide currencies (Table 1)

```
Year            US$                 DM (normalized)

1987       -15.45373781              -8.81156970
1988       -34.76981583              (0.22679764)
1989       -23.09606696               6.62319948
1990        -3.40714030               1.32634893
1991       -29.99294668              -4.47686200
1992       -35.50780036             -12.73315871
1993       -10.23731638              -9.27355928
1994       -55.41105144             -78.47416628
1995        -6.84158586             -14.39841047
1996       -10.44326395             -49.05905527
```

**Table 5**

Improvement (or worsening) of the prediction on the basis of the assumption of systemness in the nominal exchange data in comparison to univariate timelines, for the case of 15 EU member states and 11 EC member states.

| YEAR | 15 member states (EU) | 11 member states (EC) | REER (EC) |
|------|----------------------|----------------------|-----------|
| 1987 | -0.82607438 | -0.86647414 | (0.02511461) |
| 1988 | 1.68470963 | 1.86778215 | 1.24406229 |
| 1989 | (-0.01974935) | 0.29255241 | -0.11642926 |
| 1990 | (-0.08202988) | 0.31599596 | 0.29265731 |
| 1991 | -0.24174252 | 0.48141053 | -0.27362359 |
| 1992 | -0.70750947 | -0.05812540 | -0.09436024 |
| 1993 | -0.73487024 | -0.54354992 | 0.76952849 |
| 1994 | 2.40052927 | 0.18931945 | 0.52320784 |
| 1995 | (-0.09895706) | (-0.05664408) | 0.41080807 |
| 1996 | 2.38339315 | 2.09119657 | 1.15037528 |

**Table 6**

Improvement of the prediction on the basis of the Markov assumption for 10 European currencies in relation to the German Mark

**a. based on nominal exchange rates**

| COUNTRY | 1987 | 1988 | 1989 | 1990 | 1991 | 1992 | 1993 | 1994 | 1995 | 1996 |
|---|---|---|---|---|---|---|---|---|---|---|
| BELGIUM/LUX | 6.21 | 8.19 | 3.25 | -0.74 | 1.85 | 2.15 | 1.55 | 5.73 | 5.77 | 6.19 |
| DENMARK | 5.38 | 5.86 | 2.35 | -1.78 | 0.52 | 1.51 | (0.52) | 5.86 | 4.75 | 5.39 |
| FRANCE | 4.17 | 3.03 | -1.04 | -1.24 | 1.40 | 0.64 | 0.79 | 6.15 | 4.97 | 3.88 |
| IRELAND | 2.46 | -1.91 | -3.42 | -0.38 | 0.79 | 0.89 | 0.89 | 4.81 | 4.03 | 1.54 |
| ITALY | -1.25 | 1.77 | -1.56 | -0.94 | 0.88 | -1.26 | -3.37 | -14.60 | -2.07 | -10.72 |
| NETHERLANDS | 7.26 | 10.47 | 4.78 | (-0.08) | 0.87 | 0.82 | 1.24 | 7.04 | 5.69 | 5.32 |
| PORTUGAL | -14.09 | -13.14 | -9.35 | -5.19 | -5.35 | -3.25 | 2.95 | 1.85 | -7.07 | (0.01) |
| SPAIN | -3.91 | -3.63 | 2.34 | 6.90 | 5.47 | 1.25 | (-0.94) | -10.86 | -13.72 | -4.67 |
| UK | -7.09 | -8.77 | 2.94 | 3.76 | -5.95 | -2.79 | -4.18 | -5.78 | -2.28 | -4.85 |
| EC | -0.87 | 1.87 | 0.29 | 0.32 | 0.48 | -0.06 | -0.54 | 0.19 | (0.06) | 2.09 |

**b. based on real exchange rates**

| COUNTRY | 1987 | 1988 | 1989 | 1990 | 1991 | 1992 | 1993 | 1994 | 1995 | 1996 |
|---|---|---|---|---|---|---|---|---|---|---|
| BELGIUM | -1.81 | -1.75 | 0.72 | 3.01 | 1.06 | 0.94 | (0.27) | -0.71 | -3.71 | -3.29 |
| DENMARK | -2.25 | -4.28 | -1.90 | 1.81 | 1.19 | 1.42 | 3.26 | -1.02 | -1.28 | -2.15 |
| FRANCE | -1.57 | (0.02) | 1.76 | 3.22 | 1.96 | 2.50 | 2.88 | -1.86 | -1.86 | -0.82 |
| IRELAND | -2.70 | -0.64 | 4.48 | 4.16 | 1.36 | 2.03 | 1.68 | -1.98 | -1.48 | -0.14 |
| ITALY | -1.30 | -3.50 | -0.70 | -0.96 | -1.96 | -0.44 | (-1.05) | -3.79 | -2.64 | 6.49 |
| LUXEMBOURG | 0.54 | 3.20 | 1.72 | 0.74 | 1.28 | 0.70 | 1.44 | -0.48 | -2.55 | (-0.16) |
| NETHERLANDS | 1.07 | -1.99 | 1.14 | 5.00 | 4.40 | 3.15 | 1.72 | -2.57 | -3.11 | -3.08 |
| PORTUGAL | 2.91 | 4.85 | 1.50 | -3.64 | -3.81 | -7.50 | -10.64 | -4.12 | 1.86 | -0.76 |
| SPAIN | -2.11 | -3.67 | -4.16 | -7.45 | -6.11 | -1.86 | 0.96 | 7.45 | 9.42 | 2.50 |
| UK | 7.24 | 9.02 | -4.68 | -5.61 | (0.36) | -1.04 | (0.25) | 9.60 | 5.76 | 2.58 |
| EC | (0.03) | 1.24 | -0.12 | 0.29 | -0.27 | -0.09 | 0.77 | 0.52 | 0.41 | 1.15 |

**APPENDIX I**

**The prediction of the next value in a time-series**

To arrive at an information theoretical equivalent of conventional univariate time series analysis—which we will further generalize to the multivariate case—we need to transform the time series of data between the year *m* and the year *n* into a relative frequency distribution, using:

$$P_i = F_i / \Sigma_{i=m}^{n} F_i$$

The distribution may be visualized as follows:

year ->

| $P_m$ | $P_{m+1}$ | $P_{m+2}$ | | | | $P_{n-2}$ | $P_{n-1}$ | $P_n$ |

In order to extend this series to the year *n+1*, we compare the distribution in the former series (*m, m+1, ...., n-1, n*) with the distribution in the series (*m+1, m+2, ...., n, n+1*):

year ->

| $P_m$ | $P_{m+1}$ | $P_{m+2}$ | | | | $P_{n-2}$ | $P_{n-1}$ | $P_n$ | |
| | $Q_{m+1}$ | $Q_{m+2}$ | | | | $Q_{n-2}$ | $Q_{n-1}$ | $Q_n$ | $Q_{n+1}$ |

It can be shown[1] that, if we have a system of mutually exclusive events, $E_1,....,E_n$ with prior probabilities

---

[1] Cf. Theil (1972), Chapter V, in which he elaborates the theory of Markov chains in relation to probabilistic entropy measures.

$p_1,\ldots,p_n$, then the expected information value of the message which transforms the prior probabilities into the posterior probabilities $q_1,\ldots,q_n$ is given by the following formula:

$$I(q:p) = \sum_i q_i \log(q_i / p_i)$$

Without any further information, we have no reason to expect the distribution to change when it is shifted forward by one step. Our best guess is, therefore, to assume a stationary distribution.[2] Note that this assumption is much weaker than the assumption of linearity implied in ordinary least square regression analysis; in fact, we are not required to make any further assumptions about the functional form of the conditional expectation function.

Since, in the case of forecasting, the data for all years are given except for the year $n+1$, the best prediction for $Q_{n+1}$ must be based on the addition of $\Delta I = 0$ for the $(n+1)$th year, implying $Q_{n+1} = P_n$.

$Q_{n+1}$ and $P_n$ are defined in terms of the series data as follows:

$$P_n = F_n / \sum_{i=m}^{n} F_i$$

$$Q_{n+1} = F_{n+1} / \sum_{i=m+1}^{n+1} F_i$$

Thus, $Q_{n+1} = P_n$ implies:

$$F_n / \sum_{i=m}^{n} F_i = F_{n+1} / \sum_{i=m+1}^{n+1} F_i$$

---

[2] *I* is an inverse measure of the quality of the prediction. If all the *q*'s are equal to their respective *p*'s, the two distributions are identical, and hence the prediction is perfect: in this case *I* vanishes. It can also be shown that in all other cases *I* is positive; this fact corresponds to our intuitive notion that any change implies a message with additional information. (*I* may also be regarded as a measure of dissimilarity between distributions.)

However, obviously:

$$\sum_{i=m+1}^{n+1} F_i = \left(\sum_{i=m}^{n} F_i\right) - F_m + F_{n+1}$$

which implies, by substitution:

$$\frac{F_{n+1}}{\sum_{m}^{n} F_i - F_m + F_{n+1}} = \frac{F_n}{\sum_{m}^{n} F_i}$$

from which it can be derived that:

$$F_{n+1} = \left\{ \frac{\left(\sum_{m}^{n} F_i\right) - F_m}{\left(\sum_{m}^{n} F_i\right) - F_n} \right\} * F_n$$

We have now written the value of the indicator *F* for the year *n+1* as a function of the value of that same indicator in the previous years of the series. The coefficient is the sum of the time series minus the value for the first year of the series divided by the same sum minus the value of the last year.

Instead of the values of $F_i$ one could use the moving averages during a certain number of years in order to smooth out disturbances in the values for various years.[3] In fact, this gives an intuitive basis for the idea of

---

[3] Alternatively, depending on the research question, one may follow Theil (1966: 325-7) who sought to improve the predictions based on information theory by weighting more recent data linearly by using the following formula:

$$\sum F_i' = \frac{N * F_n + (N-1) * F_{n-1} + (N-2) * F_{n-2} + \dots + F_1}{\frac{1}{2} N * (N+1)}$$

potential interaction between autoregression and the moving average in time series analysis. Since the formula contains the $\Sigma_i$ for most of the series both in the numerator and in the denominator, the weighted average already functions as a momentum which cancels out some of the disturbances. In a moving average process, a disturbance affects the system for a finite number of periods (the order of the moving average) and then abruptly ceases to affect it.

However, whether we compute with raw data or with moving averages, the question remains as to how many years to include in the time series, i.e., how to determine *m* in the above formulas. Different starting years will result in different factors. If one does not want to take this decision upon visual inspection of the graphs, one can program this problem into the computer: let *m* run from 1 to *n*, and compute, for each *m*, the prediction of $F_{n+1}$, $Q_{n+1}$, and subsequently *I* as a function of all the *q*'s and *p*'s involved in this run. The *m* which leads to the lowest *I*, and correspondingly the $F_{n+1}$ derived from that *m*, can be considered as the best prediction we are able to make on the basis of this series.

**Generalization to the multi-variate case**

Analogously to the arguments presented above, we can develop the following figures for the multi-variate prediction:

---

The assumption of equality for $Q_{n+1}$ and $P_n$ in order to estimate next year's share can be maintained; however, the computation of relative frequencies and predictions becomes a bit more complex.

year ->

```
┌─────┬──────┬──────┐  ┌─ ─ ┬ ─ ─┐  ┌──────┬──────┬─────┐
│P₁,m │P₁,m+1│P₁,m+2│  │    │    │  │P₁,n-2│P₁,n-1│P₁,n │
├─────┼──────┼──────┤  ├─ ─ ┼ ─ ─┤  ├──────┼──────┼─────┤
│P₂,m │P₂,m+1│P₂,m+2│  │    │    │  │P₂,n-2│P₂,n-1│P₂,n │
├─────┼──────┼──────┤  ├─ ─ ┼ ─ ─┤  ├──────┼──────┼─────┤
│     │      │      │  │    │    │  │      │      │     │
├─────┼──────┼──────┤  ├─ ─ ┼ ─ ─┤  ├──────┼──────┼─────┤
│Pj,m │Pj,m+1│Pj,m+2│  │    │    │  │Pj,n-2│Pj,n-1│Pj,n │
└─────┴──────┴──────┘  └─ ─ ┴ ─ ─┘  └──────┴──────┴─────┘
```

```
      ┌──────┬──────┐  ┌─ ─ ┬ ─ ─┐  ┌──────┬──────┬─────┬──────┐
      │Q₁,m+1│Q₁,m+2│  │    │    │  │Q₁,n-2│Q₁,n-1│Q₁,n │Q₁,n+1│
      ├──────┼──────┤  ├─ ─ ┼ ─ ─┤  ├──────┼──────┼─────┼──────┤
      │Q₂,m+1│Q₂,m+2│  │    │    │  │Q₂,n-2│Q₂,n-1│Q₂,n │Q₂,n+1│
      ├──────┼──────┤  ├─ ─ ┼ ─ ─┤  ├──────┼──────┼─────┼──────┤
      │      │      │  │    │    │  │      │      │     │      │
      ├──────┼──────┤  ├─ ─ ┼ ─ ─┤  ├──────┼──────┼─────┼──────┤
      │Qj,m+1│Qj,m+2│  │    │    │  │Qj,n-2│Qj,n-1│Qj,n │Qj,n+1│
      └──────┴──────┘  └─ ─ ┴ ─ ─┘  └──────┴──────┴─────┴──────┘
```

As before, the best prediction for $Q_{j,n+1}$ is obtained when no substantive information is added on the basis of *a priori* reasoning. Since this occurs if and only if $Q_{j,n+1} = P_{j,n}$, we have:

$$\frac{F_{j,n+1}}{\sum_{i=m+1}^{n+1} \sum_j F_{ij}} = \frac{F_{jn}}{\sum_{i=m}^{n} \sum_j F_{ij}}$$

However, noting that

$$\sum_{i=m+1}^{n+1} \sum_j F_{ij} = \sum_j \left\{ \sum_{i=m}^{n} F_{ij} - F_{mj} + F_{n+1,j} \right\}$$

we have

$$F_{j,n+1} = F_{jn} * \left\{ \frac{\sum_{i=m}^{n} \sum_j F_{ij} - \sum_j F_{mj} + \sum_j F_{n+1,j}}{\sum_{i=m}^{n} \sum_j F_{ij}} \right\}$$

$$F_{j,n+1} = F_{jn} * \left\{ \frac{\text{Grandsum}_{mn} - \text{Columnsum}_m + \text{Columnsum}_{n+1}}{\text{Grandsum}_{mn}} \right\}$$

Since the columnn sum for the year *n+1* is a normalization factor,[4] the term on the right hand is a constant, from which may conclude that the best prediction for next year's distribution will always be the current distribution. This is also called the Markov property.

In summary, one can now make two optimal forecasts: one on the basis of the values of each individual element of the system, and another for the system as a whole, i.e., on the basis of the system's distribution in the previous year. Since the information theoretical approach is thus essentially nonparametric, its requirements with respect to the number of observations and the measurement scale are much weaker than those of conventional ARIMA time series models or cross section regression analysis.[5]

---

[4] The prediction of the value of this sum may, for example, be calculated univariately on the basis of the time series for the column sums by using methods from the previous section.

[5] On the advantage of the nonparametric character of information theoretical measures, see: Theil (1966 and 1967); Krippendorff (1986); Leydesdorff (1995). For a general introduction in using information theory, see: Theil (1972).